\documentclass[onecolumn,preprintnumbers,amsmath,aps,
]{revtex4-1}
\usepackage{graphicx}
\usepackage{dcolumn}
\usepackage{bigints}
\usepackage{bm}
\usepackage{subfigure}
\usepackage{color}
\usepackage{microtype}
\usepackage{amsmath}
\usepackage{mathtools}
\usepackage{amssymb}
\begin{document}
\def\eq#1{(\ref{#1})}
\def\fig#1{figure\hspace{1mm}\ref{#1}}
\def\tab#1{table\hspace{1mm}\ref{#1}}
\title{Cosmological aspects of the unimodular-mimetic $f(\mathcal{G})$ gravity}

\author{Adam Z. Kaczmarek}\email{a.kaczmarek@doktorant.ujd.edu.pl}
\author{Dominik Szcz{\c{e}}{\'s}niak}
\affiliation{Department of Theoretical Physics, Faculty of Science and Technology, Jan D{\l}ugosz University in Cz{\c{e}}stochowa, 13/15 Armii Krajowej Ave., 42200 Cz{\c{e}}stochowa, Poland}

\date{\today} 
\begin{abstract}

In this work we introduce and study the unimodular-mimetic $f(\mathcal{G})$ gravity, where unimodular and mimetic constraints are incorporated through corresponding Lagrange multipliers. We present field equations governing this theory and discuss their main properties. By using the reconstruction scheme, we obtain quadratic unimodular-mimetic $f(\mathcal{G})=A\mathcal{G}^2$ gravity capable of describing hybrid expansion law and the power law evolution. Furthermore, we employ an inverted reconstruction technique in order to derive specific $f(\mathcal{G})$ function that reproduces the Hubble rate of symmetric bounce. The unimodular-mimetic $f(\mathcal{G})=A\mathcal{G}^2$ is also shown to be compatible with the BICEP2/Keck and Planck data. To this end, we incorporate updated constraints on the scalar-to-tensor ratio and spectral index, utilizing a perfect fluid approach to the slow-roll parameters. Through an analysis of that kind, we demonstrate that the theoretical framework presented here can indeed characterize inflation that agrees with the observational findings. Consequently, the introduced extension appears to have potential to describe and encompass a wide spectrum of cosmological models.

\end{abstract}

\maketitle

Despite obvious success of the general relativity (GR) in the early 20th century, there have been multiple attempts to extend it. Arthur Eddington, Hermann Weyl and Kaluza-Klein were first in this field \cite{eddington1924,weyl1918,Klein1926,kaluza2018}. Later, Carl Brans and Robert Dicke developed the scalar-tensor theory, influenced by the Mach's principle \cite{jordan1955,brans1961}. These attempts were supplemented by the Lovelock's theorem, which provided other options for modifying gravity such as going beyond four dimensions, adding new fields or derivatives, consideration of nonlocal theories, and more \cite{lovelock1971}. From a slightly different perspective, Sakharov and Stelle shown also usefulness of the higher-order terms in the quantum gravity regime \cite{Sakharov_1991,stelle1977}. Yet still, the modified gravity was not widely discussed in the physics community until the advent of new ideas coming from cosmology.

In the 1980s, Starobinsky, Guth, and Linde introduced the concept of inflation, which described early exponential expansion of the Universe \cite{starobinsky1980,guth1981,LINDE1982,linde2007,clifton2012}. They recognized that the standard GR fell short in providing satisfactory explanation for key cosmological issues such as the flatness and horizon problems, the generation of primordial density fluctuations, and the extremely rapid (quasi-exponential) early expansion of the Universe. To address these challenges, they proposed alterations to Einstein's original formulation, acknowledging the need for an extended framework beyond the standard GR. In this regard, Guth introduced scalar field coupled to gravity, called inflaton, while Starobinsky proposed quantum corrections to the Einstein-Hilbert action \cite{guth1981,starobinsky1980}. This marked a crucial step in the development of theoretical physics, associated with extensions of the GR, by paving the route for novel prospects and conclusions \cite{linde2007,clifton2012}. It turned out that modification of the GR was not solely rooted in conceptual dissatisfaction with the original formulation. Instead, it emerged as a response to the challenges and shortcomings of the Einsteinian framework. The further limitations of the GR became even more evident later on, with the appearance of the Dark Universe concept. At present, there is substantial evidence that almost $68 \%$ of the Universe is filled with the dark energy that drives the accelerated expansion of the Universe, while approximately $27\%$ of the Universe is composed of the dark matter found in the galaxy clusters that is responsible for rotation of galaxies \cite{copeland2006,Garrett2010}. It means that about $95\%$ of the Universe is invisible and hence challenging to study because the dark sector does not interact electromagnetically \cite{capoziello2010,huterer2017}. These peculiar components and issues are assumed to exist if the original formulation of the general relativity is entirely accurate. Serious scepticism led to attempts to modify GR at large scales in order to better explain this mysterious behaviour \cite{clifton2012,lambiase2015}.

In the context of the above, various proposals of modified gravity have been introduced over the years, including additional curvature invariants, coupling with scalar and vector components or non-trivial interactions between matter and curvature in the action principle \cite{clifton2012,Shankaranarayanan2022}. One common approach is the $f(R)$ gravity, where the standard GR action is generalized to an arbitrary function of the Ricci scalar ($R \rightarrow f(R)$). Other invariants, such as the Gauss-Bonnet (G-B) term, have been included in the gravity's action as found within the $f(\mathcal{G})$ gravity \cite{lambiase2015}. Expansions of the GR can also involve scalar-geometry or the scalar-matter couplings. These extensions introduce terms in the gravitational action that depend on geometry, scalar fields, matter terms or the trace of the energy-momentum tensor. Examples include the Brans-Dicke theory and generalizations such as the $f(\varphi, R)$ and $f(\varphi, \mathcal{G})$ gravity \cite{brans1961,defelice2010,capozziello2011,bahamonde2015,malik2020}. The matter-geometry couplings, such as the $f(R, T)$ and $f(R, T, R_{\mu\nu}T^{\mu\nu})$ models, have also been extensively studied, often with further extensions \cite{harko2011,harko2014b,sharif2021,sardar2023,kaczmarek2020}. Note that these different approaches sometimes overlap, leading to a new results, perspectives, and conclusions \cite{cotsakis2022}. However, one have to be simultaneously wary when attempting to modify the standard GR theory. The resulting equations and models are often complex and hard to solve, with new solutions becoming relatively difficult to obtain \cite{nojiri2019}. 

Here, it is instructive to note that the modified gravity is not only limited to the dark energy problem and the associated accelerated expanse of the Universe, but also relates to other important aspects and issues \cite{clifton2012}. In particular, an interesting approach known as the mimetic theory, being introduced by Chamseddine and Mukhanov, isolates the conformal degree of freedom of gravity \cite{chamseddine2013,chamseddine2014}. This allows the mimetic field to {\it mimic} the behavior of the dark matter as well. Moreover, the extended gravity, that includes mimetic theory, plays an important role in understanding early stages of the Universe, particularly in the context of inflation and related topics \cite{clifton2012,leon2015,nojrii2017,sebastiani2017,cardenas2021,farsi2022,baffou2023}. As a result, the mimetic gravity has been successful in yielding viable models for both inflation and bounce scenarios, providing an interesting connection between the mimetic gravity and the loop quantum cosmology \cite{Langlois2017,chamseddine2014,chamseddine2019,noori2023}. These results suggest that the mimetic theory may have fundamental significance. This is additionally reinforced by the fact that the mimetic model has been extensively studied and combined with other modifications of gravity including higher-order terms, the unimodular gravity and the non-minimal couplings \cite{lambiase2015,nojiri2016,sebastiani2017,baffou2017,zhong2018,kaczmarek2021,gashti2022,bhattacharjee2022}. In this context, the unimodular approach may be of particular interest, as a modification of general relativity introduced by Einstein himself for solving the cosmological constant issue. While classically both unimodular and standard formulation of the GR are compatible, at the quantum level they are distinct from each other, leading to the different characteristics and new prospects \cite{Alvarez2005,smolin2011,alvarez2015,carballo2022}. In principle, the integration of mimetic and unimodular disciplines offers a potential solution to both the dark matter and the dynamic cosmological constant issues through a concise geometrical framework. Furthermore, the inclusion of additional geometric dependencies, such as those found in the $f(R)$ theories, enhances the flexibility and adaptability of the resulting frameworks \cite{odintsov2016}. Note, that from a technical side, unimodular and mimetic gravity can be induced at the level of theory's action by using Lagrange multipliers \cite{Astashenok2015,odintsov2016,nojiri2017b}.

In the present paper, we follow the above research trend and propose general unimodular-mimetic $f(\mathcal{G})$ gravity. This is motivated not only by the interesting and desired properties of both mimetic and unimodular theories but also the flexibility of higher-order gravities (such as the $f(R)$ and $f(\mathcal{G})$ theories) \cite{Astashenok2015,odintsov2015a,odintsov2016}. In details, after introduction of the field equations and the main properties of the new theory, we employ reconstruction scheme to derive models capable of embodying specific cosmological scenarios. Subsequently, we delve into the exploration of an inflationary model by examining its main characteristics and assessing consistency with the latest data given by ESA Planck and BICEP/Keck experiments within the perfect fluid approach to the slow-roll evolution. Next, via the reconstruction method, we obtain mimetic and unimodular Lagrange multipliers as well as the mimetic-potential responsible for inflation in the context of $f(\mathcal{G})=A\mathcal{G}^2$ gravity. Our paper concludes with a summary and discussion of future prospects and opportunities that this model unveils.

\section{Theoretical Framework}

In the mimetic gravity, the Einstein-Hilbert metric $g_{\mu\nu}$ can be written in terms of the auxiliary metric tensor $\hat{g}_{\mu\nu}$ and the auxiliary scalar field $\phi$ \cite{chamseddine2013,lambiase2015}: 
\begin{align}
    g_{\mu\nu}=-\hat{g}^{\alpha\beta}\partial_{\alpha}\phi\partial_{\beta}\phi \hat{g}_{\mu\nu}.
    \label{eq1}
\end{align}
In this manner, the original metric $g_{\mu\nu}$ can be re-expressed with the new degree of freedom linked to the invariance under conformal transformations of the metric $\hat{g}_{\mu\nu}$. Consistency of the Eq.(\ref{eq1}) leads to the mimetic constraint equation imposed on the gradient of the scalar field $\phi$ \cite{chamseddine2013}:
\begin{align}
    g^{\mu\nu}(\phi,\hat{g}_{\mu\nu})\partial_{\mu}\phi\partial_{\nu}\phi=-1.
        \label{eq2}
\end{align}
The detailed introduction to the mimetic theory can be found in the review \cite{sebastiani2017}.

On the other hand, the main idea in unimodular gravity states that while components of the spacetime metric are dynamical, the determinant of the metric itself remains fixed. This means that the determinant of the metric tensor fulfils:
\begin{align}
    \sqrt{-g}=1,
\label{eq3}
\end{align}
where  constant parameter is fixed and equal to one. In that manner, constant $\Lambda$ emerges from the traceless part of the Einstein's equations \cite{lambiase2015}.

In order to impose given constraints, we use well-known Langrange multiplier formalism, extensively used in the mimetic and unimodular theories \cite{odintsov2015a,nojiri2016,mansoori2021,odintsov2016,cardenas2021}.
In the unimodular-mimetic regime, we need to introduce two Lagrange multipliers corresponding to the unimodular ($\lambda$) and mimetic ($\eta$) constraints. Hence, for the unimodular-mimetic $f(\mathcal{G})$ gravity we have:
\begin{align}
S=\int \text{d}^4x\Big[\sqrt{-g}\big\{R+ f(\mathcal{G})-U(\phi)+\eta(g^{\mu\nu}\partial_\mu\phi\partial_\nu\phi+1)-\lambda \} +\lambda \Big]
\label{eq4}
\end{align}
where $f(\mathcal{G})$ is an analytical function of the Gauss-Bonnet term $\mathcal{G}=R^{\quad\quad\mu\nu\alpha\beta}_{\mu\nu\alpha\beta}-4R^{\mu\nu}_{\quad \mu\nu}+R^2$ and $U(\phi)$ denotes the scalar (mimetic) potential. We also note that the actions involving scalar field constrained by Lagrange multiplier can be studied without referring to the mimetic theory, see for instance \cite{capoziello2010,Gao011,nojiri2017b,nojiri2019b}.  
Further in this paper, the vacuum spacetime is assumed, hence the matter Lagrangian density is absent in the action (\ref{eq4}). This is to say, every result obtained in our work holds without invoking any special form of matter or fluid \cite{lambiase2015}.

Now, variation  of the  action  (\ref{eq1}) with respect to the components of the metric tensor $g^{\mu\nu}$ gives the field equations:
\begin{align}\nonumber
&R_{\mu\nu} -\frac{1}{2}g_{\mu\nu}\big(R+ f(\mathcal{G})-U(\phi)+\eta(g^{\mu\nu}\partial_\mu\phi\partial_\nu\phi+1)-\lambda\big)+\big(2RR_{\mu\nu}-4R^\lambda_\mu R_{\lambda\nu}-4R_{\mu\alpha\nu\beta}R^{\alpha\beta}+2R^{\lambda\ \alpha\beta}_{\ \mu}R_{\nu\lambda\beta\alpha}\big) f_\mathcal{G}
\\ \nonumber
 &+\big(2Rg_{\mu\nu}\Box-2R\nabla_\mu\nabla_\nu-4g_{\mu\nu}R^{\alpha\beta}\nabla_\alpha\nabla_\beta-4R_{\mu\nu}\Box
+4R^\lambda_\mu\nabla_\nu\nabla_\lambda+4R^\lambda_\nu\nabla_\mu\nabla_\lambda+4R_{\mu\alpha\nu\beta}\nabla^\alpha\nabla^\beta\big)f_\mathcal{G}\\ 
&+\eta\partial_\mu\phi\partial_\nu\phi=0,
\label{eq5}
\end{align}
where $f_\mathcal{G}=\frac{df(\mathcal{G})}{d\mathcal{G}}$ is the functional derivative taken with respect to the G-B term.

Variation of the action with respect to the auxiliary scalar field $\phi$ leads to the following scalar equation:
\begin{align}
    2\nabla^{\mu}(\eta \partial_{\mu}\phi)+U'(\phi)=0,
        \label{eq6}
\end{align}
where prime stands for differentiation with respect to the scalar field ($U'(\phi)=\frac{\text{d}U(\phi)}{\text{d}\phi}$).
It is important to remark, that variation of the action with respect to either $\lambda$ or $\eta$ returns unimodular or mimetic constraints, respectively.

Continuing our work, we will assume in this study that the geometry is described by isotropic and homogeneous FLRW spacetime. The line element for this spacetime takes the following form:
\begin{align}
    \text{d}s^2=-\text{d}t^2+a^2(t)(\text{d}x^2+\text{d}y^2+\text{d}z^2)
    \label{eq7}
\end{align}
in the usual Cartesian coordinates $(t,x,y,z)$. We also assume that scalar field depends only on time, {\it i.e.} $\phi=\phi(t)$. The corresponding Gauss-Bonnet term is:
\begin{align}
    \mathcal{G}=-24H^2(\dot{H}+H^2),
    \label{eq8}
\end{align}
where for brevity "dot" ($\dot{}$) denotes differentiation w.r.t. cosmic time $t$ and Hubble rate is given by $H=\frac{\dot{a}}{a}$.
Then, corresponding unimodular-mimetic $f(\mathcal{G})$ gravity equations are:
\begin{align}
3H^2+\frac{1}{2}\eta(\dot{\phi}^2+1)+\frac{1}{2}\big(f(\mathcal{G})-U(\phi)-\lambda-\mathcal{G}f_{\mathcal{G}}\big)+12H^3 \dot{f}_{\mathcal{G}} =0
    \label{eq9}
\end{align}
and
\begin{align}
    -(2\dot{H}+3H^2)-\frac{1}{2}\eta(\dot{\phi}^2-1)-\frac{1}{2}\big(f(\mathcal{G})-U(\phi)-\lambda-\mathcal{G}f_{\mathcal{G}}\big)-8H(H^2+\dot{H})\dot{f}_{\mathcal{G}}-4H^2 \Ddot{f}_{\mathcal{G}}=0,
\label{eq10}
\end{align}
i.e. for $00$th and $ii$th components respectively.
Moreover, the scalar equation take form:
\begin{align}
    6H\eta \dot{\phi}+2(\dot{\eta}\dot{\phi}+\eta\Ddot{\phi})-U'(\phi)=0.
\label{eq11}
\end{align}
Mimetic constraint for a given metric is:
\begin{align}
    \dot{\phi}^2=1,
\label{eq12}
\end{align}
which directly leads to the association of the scalar field with a cosmic time $\phi=t$. We note that, this is an important result in the mimetic gravities as in a wide class of the scalar-tensor theories, the scalar field coupled to gravity often takes more complicated forms.
By using $\phi=t$; field and scalar equations reduce to:
\begin{align}
    3H^2+\eta+\frac{1}{2}\big(f(\mathcal{G})-U(t)-\lambda-\mathcal{G}f_{\mathcal{G}}\big)+12H^3 \dot{f}_{\mathcal{G}} =0,
\label{eq13}
\end{align}
\begin{align}
    -(2\dot{H}+3H^2)-\frac{1}{2}\big(f(\mathcal{G})-U(t)-\lambda-\mathcal{G}f_{\mathcal{G}}\big)-8H(H^2+\dot{H})\dot{f}_{\mathcal{G}}-4H^2 \Ddot{f}_{\mathcal{G}}=0,
\label{eq14}
\end{align}
and:
\begin{align}
    6H\eta +2\dot{\eta}-U'(\phi=t)=0.
\label{eq15}
\end{align}
We can get rid of the  $\lambda$ by addition of $00$th and $ii$th equations:
\begin{align}
   \eta=2\dot{H}-4H(H^2-2\dot{H})\dot{f}_{\mathcal{G}}+4H^2\Ddot{f}_{\mathcal{G}}.
\label{eq16}
\end{align}
Then, from above equation one has:
\begin{align}
    \dot{\eta}=2\Ddot{H}-4H(H^2-4\dot{H})\Ddot{f}_{\mathcal{G}}   -4(3H^2\dot{H}-2\dot{H}^2-2H\Ddot{H})\dot{f}_{\mathcal{G}}+4H^2\dddot{f}_{\mathcal{G}}.
\label{eq17}
\end{align}
By combining Eq.(\ref{eq16},\ref{eq17}) we can obtain differential equation describing mimetic potential $U(t)$. Thus, potential will be given by an integral:
\begin{align}
    U(t)=\int\mathcal{E}(t)\text{d}t,
\label{eq18}
\end{align}
where
\begin{align}
  \mathcal{E}(t)=4(3H \dot{H}+\ddot{H})-8(3H^4-3H^2\dot{H}-2\dot{H}^2-2H\Ddot{H}) \dot{f_{\mathcal{G}}}+16(H^3+2H\dot{H})\ddot{f}_{\mathcal{G}}+8H^2 \dddot{f}_{\mathcal{G}}.
\label{eq19}
\end{align}
From Eq.(\ref{eq13}) one can get the expression for unimodular Lagrange multiplier $\lambda$:
\begin{align}
  \lambda(t)=  6H^2+\eta+\frac{1}{2}\big(f(\mathcal{G})-U(t)-\mathcal{G}f_{\mathcal{G}}\big)+12H^3 \dot{f}_{\mathcal{G}}.
\label{eq20}
\end{align}
Solving this, one can obtain the desired potential, after which Lagrange multipliers can be found from corresponding equations for $\eta$ and $\lambda$. This scheme will be described in details and used in the next section.
\section{Reconstruction of the cosmological scenarios}
The inherent complexity of the field equations in the modified gravity often poses a significant challenge, as obtaining exact and numerical solutions that align with observational data proves to be a hard and nontrivial task \cite{clifton2012}. However, in the reconstruction approach, the field equations are inverted to unveil the class of modified theories that give rise to a specific flat Friedmann-Lema{\^ i}tre-Robertson-Walker (FLRW) model. In other words, instead of solving FLRW equations for a given model in order to obtain scale factor, the equations are meant to reconstruct a model that satisfy a given scale factor \cite{odintsov2015a,odintsov2016,kaczmarek2020,kaczmarek2021}. Note, that ability to describe well established cosmological models in the GR modifications is desired. In this section, after specifying the function $f(\mathcal{G})$, the field equations will be solved for Lagrange multipliers $\eta$, $\lambda$ and mimetic potential $U$ that satisfy evolution of interest. At the end of this section, another viable reconstruction scheme will be presented, where obtaining unknown $f(\mathcal{G})$ function will be a main task.
\subsection{The perfect fluid}
In the present section, we assume natural and commonly used choice of the $f(\mathcal{G})$ gravity model {\it i.e.} the quadratic G-B gravity of the form \cite{zhong2018}:
\begin{align}
    f(\mathcal{G})=A \mathcal{G}^2.
\label{eq21}
\end{align}
In our initial scenario for the reconstruction method, we consider the Universe filled with a perfect fluid, that satisfy equation of state $p=\omega \rho$ for parameter $\omega$. Fluid of that kind exhibit fundamental characteristics such as isotropy and homogeneity, conservation of matter-energy as well as the macroscopic description. Such model is crucial in description of dynamics of the spacetime we live in \cite{dodelson2020}. The corresponding scale factor and the Hubble rate take form:
\begin{align}
    a(t)=a_0t^{\frac{2}{3 (\omega +1)}},\;\;\;\; H(t)=\frac{2}{3 t(\omega+1)}.
\label{eq22}
\end{align}
Usefulness of power law cosmology comes from the fact that it is suitable for describing different cosmic eras, with the example of dust dominated Universe ($\omega=0$) or radiation era ($\omega=1/3$). Moreover, models characterized by the power-law are important for specific predictions for observables such as red shift of galaxies or large-scale structures \cite{dodelson2020}.

From Eq.(\ref{eq18}), the mimetic potential reads:
\begin{align}
    U(t)=\frac{8 \big(512 A \big(252 \omega
   ^3+549 \omega ^2+362 \omega
   +69\big)-243 t^6 \omega  (\omega
   +1)^6\big)}{729 t^8 (\omega +1)^8}.
\label{eq23}
\end{align}
Inserting the potential from Eq.(\ref{eq23}) into the field equation (\ref{eq14}) leads to the expression for the Lagrange multiplier $\lambda(t)$:
\begin{align}
    \lambda(t)=-\frac{8192 A (3 \omega +1)^2}{729 t^8
   (\omega +1)^8}.
     \label{eq24}
\end{align}
Finally, the mimetic Lagrange multiplier $\lambda(t)$ obtained with the aid of Eq.(\ref{eq23}) and Eq.(\ref{eq24}) is:
\begin{align}
    \eta(t)= \frac{8192 A
   \big(63 \omega ^2+90 \omega
   +23\big)-972 t^6 (\omega +1)^6}{729
   t^8 (\omega +1)^7}.
     \label{eq25}
\end{align}
Note that the perfect fluid evolution has been obtained without inclusion of matter, as one of crucial properties of mimetic gravity and its extensions is to \textit{mimic} matter content \cite{chamseddine2013,sebastiani2017,kaczmarek2021}.
\subsection{The hybrid expansion law}
The hybrid expansion law was introduced in \cite{akarsu2014} as a product of the power-law evolutions and the flat de-Sitter cosmology. Besides describing these two phases, the hybrid scale factor allows for elegant and nice transition between two phases in cosmic history, namely deceleration to the cosmic acceleration. The corresponding scale factor and Hubble rate take the following form \cite{houndjo2012,akarsu2014}:
\begin{align}
    a(t)=a_0 \Big(\frac{t}{t_0}\Big)^{g_0}e^{g_1\big(\frac{t}{t_0}-1\big)},\;\;\; H=\frac{g_0}{t}+\frac{g_1}{t_0},
\label{eq26}
\end{align}
while again the quadratic G-B form of Eq.(\ref{eq21}) is assumed. The constants $a_0$ and $t_0$ corresponds to the present values of cosmic time and scale factor, respectively \cite{akarsu2014}.
Taking the integral given in Eq.(\ref{eq18}) with the aid of Eq.(\ref{eq17}) leads to the mimetic potential:
\begin{align}\nonumber
    U(t)&=\frac{2 g_0}{t^8 t_0^7}\Big[
    -288 A g_0^5 t_0^5 \big(-28g_1^2 t^2-20 g_1 t t_0+23t_0^2\big)+192 A g_1g_0^3 t t_0^3 \big(105g_1^3 t^3+100 g_1^2 t^2t_0-255 g_1 tt_0^2+84t_0^3\big)\\ \nonumber&+576 A g_1^2g_0^2 t^2 t_0^2 \big(28g_1^3 t^3+25 g_1^2 t^2t_0-70 g_1 tt_0^2+30t_0^3\big)+3 g_0 t^3t_0 \big(2688 A g_1^6t^3+1920 A g_1^5 t^2t_0-5280 A g_1^4 tt_0^2\\ \nonumber&+2560 A g_1^3t_0^3+t^3t_0^6\big)+192 A g_0^4t_0^4 \big(84 g_1^3t^3+75 g_1^2 t^2 t_0-150 g_1 t t_0^2+28t_0^3\big)+2 t^4\big(1152 A g_1^7 t^3+480 Ag_1^6 t^2 t_0\\ &-1152 Ag_1^5 t t_0^2+576 A g_1^4t_0^3+3 g_1 t^3t_0^6-t^2t_0^7\big)+192 A g_0^6t_0^6\big(12 g_1 t+5t_0\big)+288 A g_0^7t_0^7    \Big].
\label{eq27}
\end{align}
Then, from Eq.(\ref{eq14}) corresponding Lagrange multiplier $\lambda(t)$ is:
\begin{align}\nonumber
    \lambda (t)&=-\frac{6}{t^8 t_0^8}\Big[
    g_1^2 t^8 \big(96 A
    g_1^6-t_0^6\big)+384 A
    g_0 g_1^6 t^6 t_0 \big(4
    g_1 t-t_0\big)+192 A
    g_0^6 t_0^6 \big(28
    g_1^2 t^2-12 g_1 t
    t_0+t_0^2\big)+192 A g_0^8
    t_0^8\\ \nonumber&+38
    4 A g_0^5 g_1 t t_0^5
    \big(28 g_1^2 t^2-15 g_1 t
    t_0+2
    t_0^2\big)+384 A g_0^4
    g_1^2 t^2 t_0^4 \big(35
    g_1^2 t^2-20 g_1 t t_0+3
    t_0^2\big)-384 A g_0^7
    t_0^7 \big(t_0-4
    g_1 t\big)\\ &+192 A g_0^2
    g_1^4 t^4 t_0^2 \big(28
    g_1^2 t^2-12 g_1 t
    t_0+t_0^2\big)+38
    4 A g_0^3 g_1^3 t^3 t_0^3
    \big(28 g_1^2 t^2-15 g_1 t
    t_0+2
    t_0^2\big)\Big].
\label{eq28}
\end{align}
Finally, by using Eqs. (\ref{eq23}) and (\ref{eq24}) one can get Lagrange multiplier associated with the mimetic constraint:
\begin{align}\nonumber
    \eta(t)&=-\frac{2 g_0}{t^8 t_0^6}\Big[576 A t^2 t_0^2 g_0^2 g_1^2 \big(10 t^2 g_1^2+22 g_1 t t_0-15
   t_0^2\big)-192 A t_0^4 g_0^4 \big(-30 t^2 g_1^2-51 g_1 t
   t_0+14 t_0^2\big)+384 A g_1 t t_0^3 g_0^3 \\ \nonumber&\times\big(20 t^2
   g_1^2+42 g_1 t t_0-21 t_0^2\big)+t^4 \big(384 A t^2
   g_1^6+576 A t t_0 g_1^5-576 A t_0^2 g_1^4+t^2
   t_0^6\big)+768 A g_0 t^3 t_0 g_1^3 \\ &\times\big(3 t^2 g_1^2+6 g_1 t
   t_0-5 t_0^2\big)+2304 A t_0^5 g_0^5 \big(g_1
   t+t_0\big)+384 A t_0^6 g_0^6
    \Big].
    \label{eq29}
\end{align}
Thus, the reconstruction scheme is completed for evolution given by Eq.(\ref{eq22}) and we get unimodular-mimetic $f(\mathcal{G})$ gravity model that can reproduce hybrid Hubble rate. Interestingly, in the limit of large $t$ ($H(t)\rightarrow g_0/t_0$) the mimetic Lagrange multiplier and potential vanish, while $\lambda(t) \rightarrow \frac{6 g_1^2 \big(\text{t0}^6-96 A g_1^6\big)}{\text{t0}^8}$. Thus, the unimodular part of the model will be responsible for the de-Sitter era, together with the associated $f(\mathcal{G})$ functional. That observation elegantly corresponds to the main motivation of the unimodular gravity as a way to naturally implement cosmological constant \cite{carballo2022}.
\subsection{Reconstruction of the $f(\mathcal{G})$ function from the Lagrange multiplier and the scale factor}
The reconstruction scheme applied for the hybrid and power-law models is not the only valid option, since there exist {\it inverse} reconstruction method \cite{Astashenok2015,kaczmarek2021}. In that approach, one specify Lagrange multiplier or mimetic potential and scale factor, while the field equations are solved for the remaining unspecified functions \cite{Astashenok2015}. As a first step, one can start with rewriting Eq.(\ref{eq16}) in the form:
\begin{align}
    2\dot{H}-4H(H^2-2\dot{H})\frac{\text{d}}{\text{d}t}f_{\mathcal{G}}+4H^2\frac{\text{d}^2}{\text{d}t^2}f_{\mathcal{G}}-\eta(t)=0
\label{eq30}
\end{align}
where function $f_\mathcal{G}(t)$ is treated explicitly as a function of $t$. Solution of this equation in terms of $f_{\mathcal{G}}(t)$ reads:
\begin{align}\nonumber
    f_{\mathcal{G}}(t)&= \bigintsss^t\Big[C_1\exp \big(\int^{t_3}\frac{4H(t_1)^3-8H(t_1)\frac{\text {d}}{\text{d}t_1}H(t_1)}{4H(t_1)^2}dt_1\big)+\exp \Big(\int^{t_3}\frac{4H(t_1)^3-8H(t_1)\frac{\text {d}}{\text{d}t_1}H(t_1)}{4H(t_1)^2}dt_1\Big)\\ &\times \int ^{t_3}\frac{\exp\Big(-\int ^{t_2}\frac{4H(t_1){}^3-8H(t_1)\frac{\text {d}}{\text{d}t_1}H(t_1)}{4H(t_1)^2}dt_1\Big) \big(-\eta(t_2)-2\frac{\text {d}}{\text{d}t_2}H(t_2)\big)}{4
    H(t_2)^2}dt_2\Big]dt_3+C_2.
\label{eq31}
\end{align}
Once the $f_{\mathcal{G}}(t)$ is obtained, one can use definition of the Gauss-Bonnet term from Eq.(\ref{eq8}) to express $t$ in terms of $\mathcal{G}$. Then, upon integration with respect to $\mathcal{G}$, one gets the desired $f(\mathcal{G})$ model in the general form of:
\begin{align}
    f(\mathcal{G})=\int f_\mathcal{G}\big(t(\mathcal{G})\big) \text{d}\mathcal{G}.
\label{eq32}
\end{align}
Then, the mimetic potential and the unimodular Lagrange multiplier can be easily obtained from Eqs.(\ref{eq18}) and (\ref{eq20}), correspondingly.
As an example of this procedure, we will consider a symmetric bounce scenario, with the scale factor and the Hubble rate given as:
\begin{align}
    a(t)=e^{\alpha t^2},\;\;\;\; H(t)=2\alpha t,
\label{eq33}
\end{align}
where $\alpha$ is an arbitrary constant. Note that the scale factor is normalized in a way in which the bouncing point is located at $t=0$ with $a(0)=1$. We remark that models of that kind attempt to describe the evolution of the Universe in a way that avoids singularities and provides a more continuous and regular transition from a contracting to an expanding phase as alternatives to inflation \cite{singh2023}. It is important to emphasize, that symmetric bounce scenario needs to be integrated with other cosmological phenomena to provide a comprehensive description of primordial modes for the Hubble horizon \cite{bamba2014,kaczmarek2021}. In the present work, however, we are interested in the ability to describe cosmic evolution of that kind. The corresponding Gauss-Bonnet term takes the form:
\begin{align}
    \mathcal{G} =-192 \alpha ^3 t^2 \big(2 \alpha  t^2+1\big).
\label{eq34}
\end{align}
Inverting this relation, in order to obtain functional form of $t(\mathcal{G})$, one gets the following positive solution:
\begin{align}
    t=\sqrt{\frac{\sqrt{24 \alpha^6-\alpha^4 G}}{8\sqrt{6}\alpha^4}-\frac{1}{4\alpha}}.
\label{eq35}
\end{align}
Now, we assume that the mimetic Lagrange multiplier has the form:
\begin{align}
    \eta(t)=\mu t^2.
\label{eq36}
\end{align}
Inserting the above into the Eq.(\ref{eq15}) gives the potential $U(t)$: 
\begin{align}
    U(t) = 3 \alpha  \mu  t^4+2 \mu 
   t^2+C_1,
\label{eq37}
\end{align}
where $C_1$ is the integration constant.
Then, from Eqs. (\ref{eq30}) and (\ref{eq31}) the function $f_{\mathcal{G}}(t)$ will be given by:
\begin{align}
f_{\mathcal{G}}(t)= \frac{\sqrt{\pi } \mu G_{2,3}^{2,1}\Big(-t^2 \alpha |
\begin{array}{c}
 0,1 \\
 0,0,-\frac{1}{2} \\
\end{array}
\Big)}{128 \alpha ^3}+\frac{\sqrt{\pi } G_{2,3}^{2,1}\Big(-t^2
   \alpha |
\begin{array}{c}
 0,1 \\
 0,0,-\frac{1}{2} \\
\end{array}
\Big)}{16 \alpha }+\sqrt{\pi } \sqrt{\alpha } C_1
   \text{erfi}\big(\sqrt{\alpha } t\big)-\frac{C_1 e^{\alpha 
   t^2}}{t}+\frac{\mu  \log (t)}{32 \alpha ^3}+C_2,
\label{eq38}
\end{align}
where $\text{erfi}(x)$ and $G^{mn}_{pq}$ denote the imaginary error function and the Meijer G-function accordingly \cite{prudnikov1986}. Again, $C_1$ and $C_2$ stand for integration constants. Then, by using relationship for $t$ in terms of $\mathcal{G}$, and integrating w.r.t $\mathcal{G}$ final form is obtained. Due to the complexity of the calculations, the $f(\mathcal{G})$ function is given in the appendix. By combining functional form with the Lagrange multiplier $\eta(t)$ and potential $U(t)$, the unimodular Lagrange multiplier $\lambda(t)$ can be obtained after some work. Again, the full form is given in the appendix. 

The section ends with a short remark. Instead of assuming one of the Lagrange multipliers the mimetic potential $U(t)$ can be specified from the Eq.(\ref{eq15}). This leads to the following solution for the mimetic Lagrange multiplier $\eta(t)$:
\begin{align}
    \eta (t)=\exp\Big(\int ^t-3H(t_1)dt_1\Big)\Big(\int ^t\frac{1}{2} \exp\big(-\int ^{t_2}-3H(t_1)dt_1\big)\frac{\text{d}}{\text{d}t}U(t_2)dt_2+C\Big),
\label{eq39}
\end{align}
for the integration constant $C$.

Then, from the above equations, function $f_\mathcal{G}(t)$ as well as the $f(\mathcal{G})$ and the multiplier $\lambda(t)$ can be obtained. Hence, as a summary of this section, the unimodular-mimetic $f(\mathcal{G})$ gravity models can be reconstructed by using various approaches, once suitable assumptions has been made. Complexity of the reconstruction method also varies and depends on the method, as previously elaborated in \cite{Astashenok2015} for the mimetic $f(\mathcal{G})$ gravity.
\section{The slow-roll inflation from the perfect fluid approach}
When discussing the inflationary scenario, it is useful to employ the concept of $e$-foldings ($N$) rather than cosmic time ($t$). The number of $e$-foldings represent the measure of times during which the Universe expanded by the factor $e$ during inflation \cite{linde2007}. The relationship between the scale factor and the e-folding number $N$ can be expressed explicitly as $e^{N}=\frac{a}{a_0}$, where $a_0$ denotes the initial value of the scale factor at the beginning of the inflationary phase. To simplify matters, we list the transformation rules for time derivatives with respect to the $e$-foldings as follows:
\begin{align}
    &\frac{d}{\text{d}t}=H(N) \frac{d}{dN},\;\;\;\; \frac{d^2}{\text{d}t^2}=H^2(N)\frac{d^2}{dN^2}+H(N)H'(N)\frac{d}{dN}.
\label{eq40}
\end{align}
Note that the prime symbol ($'$) denotes the derivative with respect to the $e$-foldings number $\frac{dH(N)}{dN}$.

To investigate the slow-roll indices, we employ the perfect fluid approach developed in the study of Bamba et al.\cite{bamba2014}. In this approach, the additional terms in the gravitational action Eq.(\ref{eq1}) can be interpreted as a perfect fluid. To accomplish this, we introduce extra contributions to the action in order to model this fluid, while the mimetic field is incorporated into the equations as a pressure-less fluid. In fact, the slow-roll approximation in the effective fluid description is equivalent to the standard analysis of that kind \cite{odintsov2016}. More specifically, if someone will compare derivation of the slow-roll indices in the perfect-fluid and scalar field frameworks, there will be full agreement between them. This formalism has proven successful in the analysis of other extensions of the GR theory, such as the mimetic-unimodular $f(R)$ or the mimetic $f(\mathcal{G})$ gravity \cite{odintsov2016, zhong2018}. The advantage of this formalism lies within its ability to derive spectral indices independently of the specific model.

Following this approach, the slow-roll parameters are expressed as a functions of the Hubble rate ($H$):
\begin{align}\nonumber
    \epsilon&=\frac{-H(N)}{4H'(N)}\Big(\frac{(H'(N))^2+6H'(N)H(N)+H''(N)H(N)}{H'(N)H(N)+3H^2(N)}\Big)^2,\\ 
    \eta&=-\frac{\Big(9\frac{H'(N)}{H(N)}+3\frac{H''(N)}{H(N)}+\frac{1}{2}\Big(\frac{H'(N)}{H(N)}\Big)^2-\frac{1}{2}\Big(\frac{H''(N)}{H(N)}\Big)^2+\frac{H''(N)}{H'(N)}+\frac{H'''(N)}{H'(N)}\Big)}{2\big(3+\frac{H'(N)}{H(N)}\big)}. 
    \label{eq41}
\end{align}
Then, by using the following slow-roll indices, the spectral index of primordial curvature perturbations $n_s$ and the scalar to tensor ratio $r$ for the perfect fluid framework take form:
\begin{align}
    n_s \approx 1+2\eta-6\epsilon,\;\;\;\; r=16\epsilon.
\label{eq42}
\end{align}
Confrontation with the observations can be made by determining if it is possible to satisfy recent constraints on both the $n_s$ and the $r$ from the Planck and BICEP/Keck Array data accordingly \cite{ade2016a,ade2016}:
\begin{align}
    n_s=0.968 \pm 0.006,\;\;\;\; r<0.007.
\label{eq43}
\end{align}
Recently, however, the scalar to tensor ratio $r$ has been even more tightly constrained by using data from the BICEP/Keck telescopes \cite{ade2022}:
\begin{align}
    r<0.036,
\label{eq44}
\end{align}
within the $95\%$ confidence region. Therefore, to align with the latest experimental data, it is necessary to combine both of these constraints.
\begin{figure}
    \centering
    \includegraphics[scale=1.05]{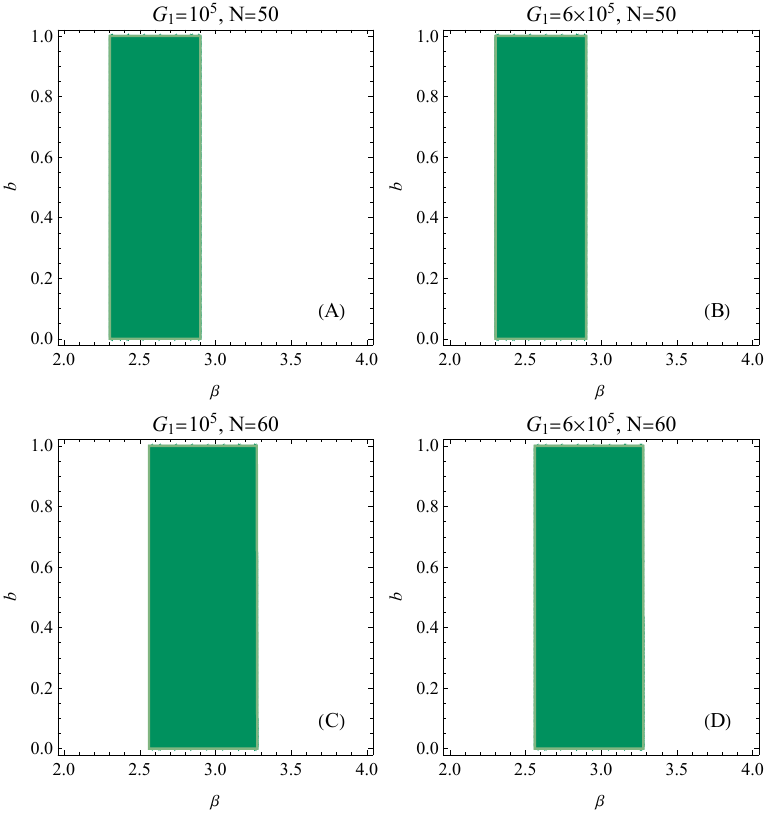}
    \caption{Regions where constraint on the scalar to tensor ratio and the spectral index values are satisfied for constant $G_0=0.00007$ and different choice of $G_1$ for different values of $\beta$ and $b$.}
    \label{fig1}
\end{figure}
First, in order to study inflation in the presented unimodular-mimetic $f\mathcal{(G)}$ gravity, the field equations are expressed in terms of the e-folding number $N$:
\begin{align}
    3H^2(N)+\eta+\frac{1}{2}(f(\mathcal{G})-U(N)-\lambda(N)-\mathcal{G}f_{\mathcal{G}})+12H^4(N)f_\mathcal{G}',
\label{eq45}
\end{align}
and
\begin{align}\nonumber
    -&(2 H(N) H'(N) + 3 H(N)^2) - \frac{1}{2} (f(\mathcal{G}) - U(N) -\lambda(N)-G f_\mathcal{G}) - 8 H(N) \big(H(N)^2 + H(N) H'(N)\big) H(N) f_\mathcal{G}' \\&- 4 H(N)^2 \Big(H(N)^2 f_\mathcal{G}'' + H(N) H'(N) f_\mathcal{G}'\Big).
\label{eq46}
\end{align}
In the manner similar to the previous approach to reconstruction method, the inflationary scenario that we wish to describe will be given by the Hubble rate \cite{odintsov2016}:
\begin{align}
    H(N)=(-G_0 N^\beta+G_1)^b,
\label{eq47}
\end{align}
with the parameters $G_0$, $G_1$, $\beta$ and $b<1$ being real numbers. Note, that parameter $b$ has to be constrained, in order to avoid Hubble rate taking negative values \cite{odintsov2016}.
The potential will be given in terms of the geometric functions and since it has very complicated nature, we refer a reader to the appendix. The associated Lagrange multipliers $\lambda(t)$ and $\eta(t)$, obtained with from Eqs.(\ref{eq45}) and (\ref{eq46}) are:
\begin{align}\nonumber
    &\lambda(N)= \frac{G_0^3 N^{3 \beta }}{N^3 \big(G_1-G_0 N^{\beta
    }\big)^3} \big(384 A b \beta  \big(28 b^2 \beta ^2-15 b \beta +2\big)
    \big(G_1-G_0 N^{\beta }\big)^{8 b}+576 A b \beta  n (25 b \beta -4) \big(G_1-G_0 N^{\beta
    }\big)^{8 b}\\ \nonumber &+N^3 \big(576 A \big(G_1-G_0 N^{\beta }\big)^{8 b}-6 \big(G_1-G_0 N^{\beta
    }\big)^{2 b}+U(N)\big)+4 b \beta  N^2 \big(G_1-G_0 N^{\beta }\big)^{2 b} \big(1056 A
    \big(G_1-G_0 N^{\beta }\big)^{6 b}-1\big)\big)\\ \nonumber &+G_0^2 G_1 N^{2 \beta } \big(-384 A b(\beta -1) \beta  (15 b \beta -\beta -4) \big(G_1-G_0 N^{\beta }\big)^{8 b}-576 A b \beta  N (25 b \beta+4 \beta -8) \big(G_1-G_0 N^{\beta }\big)^{8 b}\\ \nonumber &-3 N^3 \big(576 A \big(G_1-G_0 N^{\beta}\big)^{8 b}-6 \big(G_1-G_0 N^{\beta }\big)^{2 b}+U(N)\big)-8 b \beta  N^2 \big(G_1-G_0
    N^{\beta }\big)^{2 b} \big(1056 A \big(G_1-G_0 N^{\beta }\big)^{6 b}-1\big)\big)\\ \nonumber &-G_1^3 N^3
    \big(576 A \big(G_1-G_0 N^{\beta }\big)^{8 b}-6 \big(G_1-G_0 N^{\beta }\big)^{2
    b}+U(N)\big)+G_0 G_1^2 N^{\beta } \big(384 A b \beta  \big(\beta ^2-3 \beta +2\big)\big(G_1-G_0 N^{\beta }\big)^{8 b}\\ \nonumber &+2304 A b (\beta -1) \beta  N \big(G_1-G_0 N^{\beta
    }\big)^{8 b}+3 N^3 \big(576 A \big(G_1-G_0 N^{\beta }\big)^{8 b}-6 \big(G_1-G_0 N^{\beta
    }\big)^{2 b}+U(N)\big)\\ &+4 b \beta  N^2 \big(G_1-G_0 N^{\beta }\big)^{2 b} \big(1056 A
    \big(G_1-G_0 N^{\beta }\big)^{6 b}-1\big)\big).
\label{eq48}
\end{align}
and
\begin{align}\nonumber
    \eta(N) &=  2 b \beta  G_0 N^{\beta -3} \big(G_1-G_0
    N^{\beta }\big)^{2 b-3} \big(-G_0^2 N^{2 \beta } \big(-96 A \big(28 b^2
    \beta ^2-15 b \beta +2\big) \big(G_1-G_0 N^{\beta }\big)^{6 b}-288 A N
    (8 b \beta -1)\\ \nonumber &\times \big(G_1-G_0 N^{\beta }\big)^{6 b}+N^2 \big(384 A
    \big(G_1-G_0 N^{\beta }\big)^{6 b}+1\big)\big)-G_1^2 \big(-96
    A \big(\beta ^2-3 \beta +2\big) \big(G_1-G_0 N^{\beta }\big)^{6
    b}\\ \nonumber &-288 A (\beta -1) N \big(G_1-G_0 N^{\beta }\big)^{6 b}+N^2 \big(384 A
    \big(G_1-G_0 N^{\beta }\big)^{6 b}+1\big)\big)+2 G_0 G_1
    N^{\beta } \big(-144 A N (8 b \beta +\beta -2) \\  &\times \big(G_1-G_0 N^{\beta
    }\big)^{6 b}-48 A (\beta -1) ((15 b-1) \beta -4) \big(G_1-G_0 N^{\beta
    }\big)^{6 b}+N^2 \big(384 A \big(G_1-G_0 N^{\beta }\big)^{6b}+1\big)\big)\big)
\label{eq49}
\end{align}
accordingly. The potential obtained from the Eq.(\ref{eq18}) in terms of the $e$-foldings number gets complicated and the full form is given in the appendix (in particular, see Eq.(B1)). Thus, the inflationary model is reconstructed from the given Hubble rate and the functional form of the $f(\mathcal{G})$.

\begin{figure}
    \centering
    \includegraphics[scale=1.05]{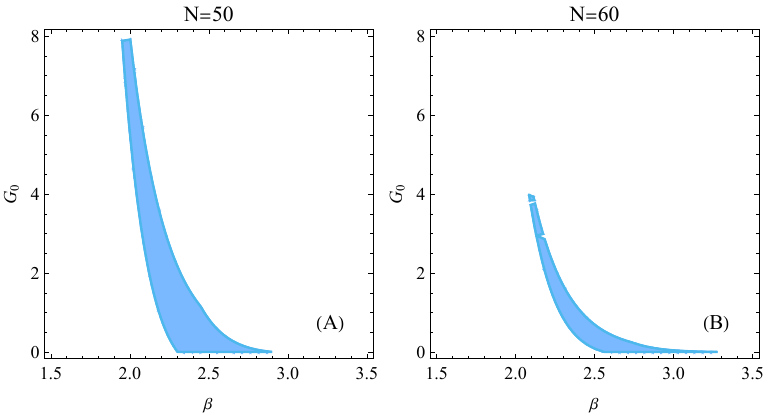}
    \caption{Regions where constraint on the scalar to tensor ratio and the spectral index values are satisfied for constant $G_1=2\times 10^5$ and different choice of $b=2$ as a function of $\beta$ and $G_0$.}
    \label{fig2}
\end{figure}

Since one of the objectives in the present paper is to reconstruct viable inflation with the example presented above, it is worth to update the relationship of the model studied previously in the context of the unimodular-mimetic $f(R)$ gravity \cite{odintsov2016}, as the new constraints on the slow-roll parameters have been imposed \cite{ade2022}. Thus, example of values of the parameters that are in agreement with the BICEP/Keck and Planck data are presented on the Figs. (\ref{fig1}) and (\ref{fig2}), for both viable $e$-foldings numbers ($N=50$ and $N=60$). Hence, for appropriate choice of the parameters associated with the inflationary Hubble rate, the unimodular-mimetic extension of the $f(\mathcal{G})$ gravity does in fact include inflation.

\section{Summary and  discussion}
In the present work, a new approach to the unimodular-mimetic gravity that incorporates higher-order Gauss-Bonnet terms has been introduced by the Lagrange multiplier approach. The unimodular-mimetic $f(\mathcal{G})$ gravity of that kind can reconstruct viable and well-established scenarios of cosmological evolution. In this regard, for the simplest functional dependence of the $f(\mathcal{G})=A\mathcal{G}^2$, an analysis has been performed by successfully obtaining the Lagrange multipliers $\eta(t)$ and $\lambda(t)$ as well as the potential $U(t)$ that correctly portray perfect-fluid cosmology, where the Hubble rate is $H(t)=2/(3t(\omega+1))$. Note that, since matter role is played by the geometry and constraints, this model serves as another good example of the mimetic approach. Specifically, the mimetic constraint has the capability to effectively emulate the presence of matter \cite{odintsov2016}. Likewise, hybrid expansion law specified by the Hubble rate $H(t)=g_0/t+g_1/t_0$ also has been reconstructed. Since the model of that form can characterize transition between matter dominated and accelerated phases, the unimodular-mimetic extension introduced in our work is capable to describe distinct stages of Universe evolution in the unified manner. Next, an inverted reconstruction procedure has been applied, once the multiplier $\eta(t)$ was specified for the symmetric bounce model. Despite the complexity of the resulting problem, it was possible to obtain a corresponding functional form of the $f(\mathcal{G})$ model. We note that the qualitative character of the obtained $f(\mathcal{G})$ function is close to the scheme used in \cite{Astashenok2015}, where only mimetic degree of freedom was studied for the Gauss-Bonnet theories. Our work ends with the study of inflation in the unimodular-mimetic $f(\mathcal{G})=A \mathcal{G}^2$ model, where the inflationary Hubble rate $ H(N)=(-G_0 N^\beta+G_1)^b$ has been reconstructed and confronted with the experimental data for the scalar to tensor ratio and the spectral index, showing that for a suitable choice of the constants, the present theory can describe inflation in a manner that is consistent with the recent observations and updated constraints. 

We also want to point out that the theory of that kind may be a more effective way to tackle cosmological issues than the standard formulation of the mimetic gravity. As previously argued by authors of \cite{Astashenok2015}, inclusion of the $f(\mathcal{G})$ in considerations provide yet another effective degree of freedom. Since in principle the model presented here describes inflation by using the slow-roll assumption, the degree of freedom related to the Gauss-Bonnet invariant should be treated as some form of the curvaton or direct modification to mimetic dark matter. By doing so, one can address large-scale structure as-well, as an alternative for the two-field mimetic models \cite{Astashenok2015,zhang2023}. As such, more detailed study of inflation and cosmological perturbations as well as the stability of possible unimodular-mimetic models has yet to be conducted. On the other hand, the unimodular degree of freedom as an effective cosmological constant may be responsible for the de-Sitter evolution, since both the mimetic Lagrange multiplier and the potential (i.e. $\eta(t)$ and $U(t)$) vanish when hybrid scale factor $H\rightarrow g_1/t_0$. Hence, in principle, the unimodular-mimetic $f(\mathcal{G})$ gravity and its relatives can describe both the dark matter and the dark energy without any matter fluids. However, more details are needed in that manner, specifically phase-system analysis as well as the thermodynamics and energy conditions of the effective energy-momentum tensor \cite{zubair2016,haghani2018}. We also acknowledge that to enhance the viability of the unimodular-mimetic $f(\mathcal{G})$ model further investigations in this aspect are required. The forthcoming research will focus on confronting the model presented here with the observational datasets, the local estimates of the Hubble constant and the general relativity tests \cite{Berti2015}. Thus, proper astrophysical tests of the unimodular-mimetic $f(\mathcal{G})$ gravity are yet to be studied.
\bibliographystyle{apsrev}
\bibliography{biblo}
\appendix
\section{Solutions for the $f(\mathcal{G})$ function}
The reconstruction of the bounce model is continued here, due to the complexity of equations. The $f_\mathcal{G}$ function in terms of the Gauss-Bonnet term reads:
\begin{align}
   f_\mathcal{G}(\mathcal{G})=& \frac{\sqrt{\pi } G_{2,3}^{2,1}\big(\alpha 
   \big(\frac{1}{4 \alpha }-\frac{\sqrt{24 \alpha
   ^6-\mathcal{G} \alpha ^4}}{8 \sqrt{6} \alpha ^4}\big)|
\begin{array}{c}
 0,1 \\
 0,0,-\frac{1}{2} \\
\end{array}
\big)}{16 \alpha }+\frac{\sqrt{\pi } \mu 
   G_{2,3}^{2,1}\big(\alpha  \big(\frac{1}{4 \alpha
   }-\frac{\sqrt{24 \alpha ^6-\mathcal{G} \alpha ^4}}{8
   \sqrt{6} \alpha ^4}\big)|
\begin{array}{c}
 0,1 \\
 0,0,-\frac{1}{2} \\
\end{array}
\big)}{128 \alpha ^3}\\\nonumber &+\sqrt{\pi } \sqrt{\alpha } C_1
   \text{erfi}\big(\sqrt{\alpha } \sqrt{\frac{\sqrt{24
   \alpha ^6-\alpha ^4 \mathcal{G}}}{8 \sqrt{6} \alpha
   ^4}-\frac{1}{4 \alpha }}\big)-\frac{C_1 e^{\alpha 
   \big(\frac{\sqrt{24 \alpha ^6-\alpha ^4 \mathcal{G}}}{8
   \sqrt{6} \alpha ^4}-\frac{1}{4 \alpha
   }\big)}}{\sqrt{\frac{\sqrt{24 \alpha ^6-\alpha ^4
   \mathcal{G}}}{8 \sqrt{6} \alpha ^4}-\frac{1}{4 \alpha
   }}}+\frac{\mu  \log \big(\sqrt{\frac{\sqrt{24 \alpha
   ^6-\alpha ^4 \mathcal{G}}}{8 \sqrt{6} \alpha
   ^4}-\frac{1}{4 \alpha }}\big)}{32 \alpha ^3}+C_2
\end{align}
Integration with respect to the $\mathcal{G}$ gives the final form of the solution:
\begin{align}
  f(\mathcal{G})=&  \sqrt{\pi } \sqrt{\alpha } c_1 \big(192 \alpha
   ^2+\mathcal{G}\big) \text{erfi}\big(\sqrt{\alpha }
   \sqrt{\frac{\sqrt{24 \alpha ^6-\alpha ^4 \mathcal{G}}}{8
   \sqrt{6} \alpha ^4}-\frac{1}{4 \alpha }}\big)+C_2 \mathcal{G}\\ \nonumber &-\frac{192
   \sqrt{\pi } \alpha ^{9/2} c_1 \sqrt{\frac{\sqrt{6}
   \sqrt{24 \alpha ^6-\alpha ^4 \mathcal{G}}-12 \alpha
   ^3}{\alpha ^4}} \text{erfi}\big(\frac{\sqrt{\sqrt{16
   \alpha ^6-\frac{2 \alpha ^4 \mathcal{G}}{3}}-4 \alpha
   ^3}}{4 \alpha ^{3/2}}\big)}{\sqrt{\sqrt{6} \sqrt{24
   \alpha ^6-\alpha ^4 \mathcal{G}}-12 \alpha
   ^3}}-\frac{\mathcal{G} \mu }{128 \alpha ^3}+\frac{\mu 
   \sqrt{36 \alpha ^6-\frac{3 \alpha ^4 \mathcal{G}}{2}}}{16
   \alpha ^4}\\ \nonumber &+\frac{3 \mu  \log \big(12 \alpha ^3-\sqrt{6}
   \sqrt{24 \alpha ^6-\alpha ^4 \mathcal{G}}\big)}{8
   \alpha }-\frac{3 \mu  \log \big(\frac{\sqrt{6} \sqrt{24
   \alpha ^6-\alpha ^4 \mathcal{G}}-12 \alpha ^3}{48 \alpha
   ^4}\big)}{8 \alpha }+\frac{\mathcal{G} \mu  \log
   \big(\frac{\sqrt{6} \sqrt{24 \alpha ^6-\alpha ^4
   \mathcal{G}}-12 \alpha ^3}{48 \alpha ^4}\big)}{64
   \alpha ^3}\\ \nonumber &+24 \alpha ^3 c_1 e^{\frac{\sqrt{4 \alpha
   ^6-\frac{\alpha ^4 \mathcal{G}}{6}}}{8 \alpha
   ^3}-\frac{1}{4}} \sqrt{\frac{3 \sqrt{6} \sqrt{24 \alpha
   ^6-\alpha ^4 \mathcal{G}}-36 \alpha ^3}{\alpha ^4}}+2 c_1
   e^{\frac{\sqrt{4 \alpha ^6-\frac{\alpha ^4
   \mathcal{G}}{6}}}{8 \alpha ^3}-\frac{1}{4}} \sqrt{48
   \alpha ^6-2 \alpha ^4 \mathcal{G}} \sqrt{\frac{\sqrt{6}
   \sqrt{24 \alpha ^6-\alpha ^4 \mathcal{G}}}{\alpha
   ^4}-\frac{12}{\alpha }} \\ \nonumber
   &-\frac{3 \sqrt{\pi } \mu  \sqrt{\alpha ^6 \big(1-4
   \big(\frac{1}{4}-\frac{\sqrt{4 \alpha ^6-\frac{\alpha ^4
   \mathcal{G}}{6}}}{8 \alpha ^3}\big)\big)^2}
   G_{3,4}^{2,2}\big(\frac{1}{4}-\frac{\sqrt{4 \alpha
   ^6-\frac{\mathcal{G} \alpha ^4}{6}}}{8 \alpha ^3}|
\begin{array}{c}
 1,1,2 \\
 1,1,0,\frac{1}{2} \\
\end{array}
\big)}{2 \alpha ^4 \big(4 \big(\frac{1}{4}-\frac{\sqrt{4
   \alpha ^6-\frac{\alpha ^4 \mathcal{G}}{6}}}{8 \alpha
   ^3}\big)-1\big)}\\ \nonumber &
   +\frac{6 \sqrt{\pi } \mu 
   \sqrt{\alpha ^6 \big(1-4 \big(\frac{1}{4}-\frac{\sqrt{4
   \alpha ^6-\frac{\alpha ^4 \mathcal{G}}{6}}}{8 \alpha
   ^3}\big)\big)^2}
   G_{3,4}^{2,2}\big(\frac{1}{4}-\frac{\sqrt{4 \alpha
   ^6-\frac{\mathcal{G} \alpha ^4}{6}}}{8 \alpha ^3}|
\begin{array}{c}
 1,2,3 \\
 2,2,0,\frac{3}{2} \\
\end{array}
\big)}{\alpha ^4 \big(4 \big(\frac{1}{4}-\frac{\sqrt{4
   \alpha ^6-\frac{\alpha ^4 \mathcal{G}}{6}}}{8 \alpha
   ^3}\big)-1\big)}\\ \nonumber &-\frac{12 \sqrt{\pi } \sqrt{\alpha ^6
   \big(1-4 \big(\frac{1}{4}-\frac{\sqrt{4 \alpha
   ^6-\frac{\alpha ^4 \mathcal{G}}{6}}}{8 \alpha
   ^3}\big)\big)^2}
   G_{3,4}^{2,2}\big(\frac{1}{4}-\frac{\sqrt{4 \alpha
   ^6-\frac{\mathcal{G} \alpha ^4}{6}}}{8 \alpha ^3}|
\begin{array}{c}
 1,1,2 \\
 1,1,0,\frac{1}{2} \\
\end{array}
\big)}{\alpha ^2 \big(4 \big(\frac{1}{4}-\frac{\sqrt{4
   \alpha ^6-\frac{\alpha ^4 \mathcal{G}}{6}}}{8 \alpha
   ^3}\big)-1\big)}\\ \nonumber &+\frac{48 \sqrt{\pi } \sqrt{\alpha ^6
   \big(1-4 \big(\frac{1}{4}-\frac{\sqrt{4 \alpha
   ^6-\frac{\alpha ^4 \mathcal{G}}{6}}}{8 \alpha
   ^3}\big)\big)^2}
   G_{3,4}^{2,2}\big(\frac{1}{4}-\frac{\sqrt{4 \alpha
   ^6-\frac{\mathcal{G} \alpha ^4}{6}}}{8 \alpha ^3}|
\begin{array}{c}
 1,2,3 \\
 2,2,0,\frac{3}{2} \\
\end{array}
\big)}{\alpha ^2 \big(4 \big(\frac{1}{4}-\frac{\sqrt{4
   \alpha ^6-\frac{\alpha ^4 \mathcal{G}}{6}}}{8 \alpha
   ^3}\big)-1\big)},
\end{align}
where in order to integrate Meijer G-function, we have made the substitution $y=\frac{1}{4}-\frac{\sqrt{4 \alpha ^6-\frac{\alpha ^4 \mathcal{G}}{6}}}{8 \alpha ^3}$ (i.e. $\mathcal{G}=-192 \big(2 \alpha ^2 y^2-\alpha ^2 y\big)$, $dy=\frac{\alpha }{96 \sqrt{4 \alpha ^6-\frac{\alpha ^4 \mathcal{G}}{6}}}d\mathcal{G}$).

Now, the unimodular Lagrange multiplier $\lambda(t)$ is equal to:
\begin{align}
 &\lambda(t)=2 \big(-\frac{96 \sqrt{\pi } \sqrt{\frac{\sqrt{6} \sqrt{24 \alpha ^6+24
   \big(16 t^4 \alpha ^4+8 t^2 \alpha ^3\big) \alpha ^4}-12 \alpha ^3}{\alpha ^4}} C_1
\text{erfi}\big(\frac{\sqrt{\sqrt{16 \alpha ^6+16 \big(16 t^4 \alpha ^4+8 t^2 \alpha ^3\big) \alpha
   ^4}-4 \alpha ^3}}{4 \alpha ^{3/2}}\big) \alpha ^{9/2}}{\sqrt{\sqrt{6} \sqrt{24 \alpha ^6+24 \big(16
   t^4 \alpha ^4+8 t^2 \alpha ^3\big) \alpha ^4}-12 \alpha ^3}}\\ \nonumber &+12 e^{\frac{\sqrt{4 \alpha ^6+4 \big(16
   t^4 \alpha ^4+8 t^2 \alpha ^3\big) \alpha ^4}}{8 \alpha ^3}-\frac{1}{4}} \sqrt{\frac{3 \sqrt{6}
   \sqrt{24 \alpha ^6+24 \big(16 t^4 \alpha ^4+8 t^2 \alpha ^3\big) \alpha ^4}-36 \alpha ^3}{\alpha
   ^4}} C_1 \alpha ^3+12 t^2 \alpha ^2\\ \nonumber&+16 t^2 \big(-\frac{e^{t^2 \alpha } \sqrt{\pi }
   \big(1-Q\big(\frac{1}{2},t^2 \alpha \big)\big) \mu }{32 \alpha ^2 \sqrt{t^2 \alpha
   }}+\frac{e^{t^2 \alpha } \sqrt{\pi } \big(1-Q\big(\frac{1}{2},t^2 \alpha \big)\big) \mu }{64 t^2
   \alpha ^3 \sqrt{t^2 \alpha }}+\frac{e^{t^2 \alpha } \sqrt{\pi } \big(1-Q\big(\frac{1}{2},t^2 \alpha
   \big)\big) \mu }{64 \alpha ^2 \big(t^2 \alpha \big)^{3/2}}-\frac{\mu }{16 t^2 \alpha ^3}\\ \nonumber&+\frac{2
   e^{t^2 \alpha } \alpha  C_1}{t}-\frac{2 e^{t^2 \alpha } C_1}{t^3}+\frac{e^{t^2 \alpha } \sqrt{\pi }
   \big(1-Q\big(\frac{1}{2},t^2 \alpha \big)\big)}{8 t^2 \alpha  \sqrt{t^2 \alpha }}-\frac{e^{t^2
   \alpha } \sqrt{\pi } \big(1-Q\big(\frac{1}{2},t^2 \alpha \big)\big)}{4 \sqrt{t^2 \alpha
   }}+\frac{e^{t^2 \alpha } \sqrt{\pi } \big(1-Q\big(\frac{1}{2},t^2 \alpha \big)\big)}{8 \big(t^2
   \alpha \big)^{3/2}}\\ \nonumber&-\frac{1}{4 t^2 \alpha }\big) \alpha ^2-\frac{3}{2} t^4 \mu  \alpha +16 t
   \big(4 t^2 \alpha ^2+2 \alpha \big) \big(-\frac{e^{t^2 \alpha } \sqrt{\pi }
   \big(1-Q\big(\frac{1}{2},t^2 \alpha \big)\big) \mu }{64 t \alpha ^3 \sqrt{t^2 \alpha }}+\frac{\mu
   }{32 t \alpha ^3}+\frac{e^{t^2 \alpha } C_1}{t^2}-\frac{e^{t^2 \alpha } \sqrt{\pi }
   \big(1-Q\big(\frac{1}{2},t^2 \alpha \big)\big)}{8 t \alpha  \sqrt{t^2 \alpha }}\big) \alpha \\ \nonumber&+4
   \alpha +\frac{1}{2} \sqrt{\pi } \big(192 \alpha ^2-24 \big(16 t^4 \alpha ^4+8 t^2 \alpha
   ^3\big)\big) C_1 \text{erfi}\big(\sqrt{\alpha } \sqrt{\frac{\sqrt{24 \alpha ^6+24 \big(16 t^4
   \alpha ^4+8 t^2 \alpha ^3\big) \alpha ^4}}{8 \sqrt{6} \alpha ^4}-\frac{1}{4 \alpha }}\big)
   \sqrt{\alpha }\\ \nonumber&-t^2 \mu +e^{\frac{\sqrt{4 \alpha ^6+4 \big(16 t^4 \alpha ^4+8 t^2 \alpha ^3\big)
   \alpha ^4}}{8 \alpha ^3}-\frac{1}{4}} \sqrt{48 \alpha ^6+48 \big(16 t^4 \alpha ^4+8 t^2 \alpha
   ^3\big) \alpha ^4} \sqrt{\frac{\sqrt{6} \sqrt{24 \alpha ^6+24 \big(16 t^4 \alpha ^4+8 t^2 \alpha
   ^3\big) \alpha ^4}}{\alpha ^4}-\frac{12}{\alpha }} C_1\\ \nonumber&-\frac{C_1}{2}-12 \big(16 t^4 \alpha ^4+8 t^2
   \alpha ^3\big) C_2+12 \big(16 t^4 \alpha ^4+8 t^2 \alpha ^3\big) \big(\sqrt{\pi } \sqrt{\alpha }
   \text{erfi}\big(t \sqrt{\alpha }\big) C_1-\frac{e^{t^2 \alpha } C_1}{t}+C_2+\frac{\mu  \log (t)}{32
   \alpha ^3}\\ \nonumber&+\frac{\sqrt{\pi } \mu  G_{2,3}^{2,1}\big(-t^2 \alpha |
\begin{array}{c}
 0,1 \\
 0,0,-\frac{1}{2} \\
\end{array}
\big)}{128 \alpha ^3}+\frac{\sqrt{\pi } G_{2,3}^{2,1}\big(-t^2 \alpha |
\begin{array}{c}
 0,1 \\
 0,0,-\frac{1}{2} \\
\end{array}
\big)}{16 \alpha }\big)+\frac{3 \mu  \log \big(12 \alpha ^3-\sqrt{6} \sqrt{24 \alpha ^6+24 \big(16
   t^4 \alpha ^4+8 t^2 \alpha ^3\big) \alpha ^4}\big)}{16 \alpha }\\ \nonumber&-\frac{3 \mu  \log
   \big(\frac{\sqrt{6} \sqrt{24 \alpha ^6+24 \big(16 t^4 \alpha ^4+8 t^2 \alpha ^3\big) \alpha ^4}-12
   \alpha ^3}{48 \alpha ^4}\big)}{16 \alpha }\\ \nonumber&-\frac{6 \sqrt{\pi } \sqrt{\alpha ^6 \big(1-4
   \big(\frac{1}{4}-\frac{\sqrt{4 \alpha ^6+4 \big(16 t^4 \alpha ^4+8 t^2 \alpha ^3\big) \alpha ^4}}{8
   \alpha ^3}\big)\big)^2} G_{3,4}^{2,2}\big(\frac{1}{4}-\frac{\sqrt{4 \alpha ^6+4 \big(16 t^4
   \alpha ^4+8 t^2 \alpha ^3\big) \alpha ^4}}{8 \alpha ^3}|
\begin{array}{c}
 1,1,2 \\
 1,1,0,\frac{1}{2} \\
\end{array}
\big)}{\big(4 \big(\frac{1}{4}-\frac{\sqrt{4 \alpha ^6+4 \big(16 t^4 \alpha ^4+8 t^2 \alpha ^3\big)
   \alpha ^4}}{8 \alpha ^3}\big)-1\big) \alpha ^2}\\ \nonumber&+\frac{24 \sqrt{\pi } \sqrt{\alpha ^6 \big(1-4
   \big(\frac{1}{4}-\frac{\sqrt{4 \alpha ^6+4 \big(16 t^4 \alpha ^4+8 t^2 \alpha ^3\big) \alpha ^4}}{8
   \alpha ^3}\big)\big)^2} G_{3,4}^{2,2}\big(\frac{1}{4}-\frac{\sqrt{4 \alpha ^6+4 \big(16 t^4
   \alpha ^4+8 t^2 \alpha ^3\big) \alpha ^4}}{8 \alpha ^3}|
\begin{array}{c}
 1,2,3 \\
 2,2,0,\frac{3}{2} \\
\end{array}
\big)}{\big(4 \big(\frac{1}{4}-\frac{\sqrt{4 \alpha ^6+4 \big(16 t^4 \alpha ^4+8 t^2 \alpha ^3\big)
   \alpha ^4}}{8 \alpha ^3}\big)-1\big) \alpha ^2}\\ \nonumber&+\frac{3 \big(16 t^4 \alpha ^4+8 t^2 \alpha
   ^3\big) \mu }{32 \alpha ^3}-\frac{3 \big(16 t^4 \alpha ^4+8 t^2 \alpha ^3\big) \mu  \log
   \big(\frac{\sqrt{6} \sqrt{24 \alpha ^6+24 \big(16 t^4 \alpha ^4+8 t^2 \alpha ^3\big) \alpha ^4}-12
   \alpha ^3}{48 \alpha ^4}\big)}{16 \alpha ^3}\\ \nonumber&+\frac{\sqrt{36 \alpha ^6+36 \big(16 t^4 \alpha ^4+8 t^2
   \alpha ^3\big) \alpha ^4} \mu }{32 \alpha ^4}\\ \nonumber&-\frac{3 \sqrt{\pi } \sqrt{\alpha ^6 \big(1-4
   \big(\frac{1}{4}-\frac{\sqrt{4 \alpha ^6+4 \big(16 t^4 \alpha ^4+8 t^2 \alpha ^3\big) \alpha ^4}}{8
   \alpha ^3}\big)\big)^2} \mu  G_{3,4}^{2,2}\big(\frac{1}{4}-\frac{\sqrt{4 \alpha ^6+4 \big(16 t^4
   \alpha ^4+8 t^2 \alpha ^3\big) \alpha ^4}}{8 \alpha ^3}|
\begin{array}{c}
 1,1,2 \\
 1,1,0,\frac{1}{2} \\
\end{array}
\big)}{4 \big(4 \big(\frac{1}{4}-\frac{\sqrt{4 \alpha ^6+4 \big(16 t^4 \alpha ^4+8 t^2 \alpha
   ^3\big) \alpha ^4}}{8 \alpha ^3}\big)-1\big) \alpha ^4}\\ \nonumber&+\frac{3 \sqrt{\pi } \sqrt{\alpha ^6
   \big(1-4 \big(\frac{1}{4}-\frac{\sqrt{4 \alpha ^6+4 \big(16 t^4 \alpha ^4+8 t^2 \alpha ^3\big)
   \alpha ^4}}{8 \alpha ^3}\big)\big)^2} \mu  G_{3,4}^{2,2}\big(\frac{1}{4}-\frac{\sqrt{4 \alpha ^6+4
   \big(16 t^4 \alpha ^4+8 t^2 \alpha ^3\big) \alpha ^4}}{8 \alpha ^3}|
\begin{array}{c}
 1,2,3 \\
 2,2,0,\frac{3}{2} \\
\end{array}
\big)}{\big(4 \big(\frac{1}{4}-\frac{\sqrt{4 \alpha ^6+4 \big(16 t^4 \alpha ^4+8 t^2 \alpha ^3\big)
   \alpha ^4}}{8 \alpha ^3}\big)-1\big) \alpha ^4}\big)
\end{align}
\section{The reconstruction of inflation}

The mimetic potential in the full form is equal to:
{\tiny  
  \setlength{\abovedisplayskip}{2pt}
  \setlength{\belowdisplayskip}{\abovedisplayskip}
  \setlength{\abovedisplayshortskip}{0pt}
  \setlength{\belowdisplayshortskip}{0pt}
\begin{align}\nonumber
   U(N)= &\frac{4}{5} \big(G_1-G_0 N^{\beta }\big)^{3 b} \times \Biggl[-2880 A \big(G_1-G_0 N^{\beta }\big)^{6 b}
   \big(1-\frac{G_0 N^{\beta }}{G_1}\big)^{-9 b}+5\\ \nonumber &+\frac{480 A b G_0^2 N^{2 \beta -1}
   \big(G_1-G_0 N^{\beta }\big)^{6 b} \beta  (17 b \beta -1) \, _2F_1\big(2-9 b,2-\frac{1}{\beta
   };3-\frac{1}{\beta };\frac{G_0 N^{\beta }}{G_1}\big) \big(1-\frac{G_0 N^{\beta
   }}{G_1}\big)^{-9 b}}{G_1^2 (2 \beta -1)}\\ \nonumber &-\frac{480 A b G_0 N^{\beta -1} \big(G_1-G_0
   N^{\beta }\big)^{6 b} \beta  \, _2F_1\big(2-9 b,\frac{\beta -1}{\beta };2-\frac{1}{\beta };\frac{G_0 N^{\beta
   }}{G_1}\big) \big(1-\frac{G_0 N^{\beta }}{G_1}\big)^{-9 b}}{G_1}\\ \nonumber &-\frac{26640 A b^2
   G_0^2 N^{2 \beta -2} \big(G_1-G_0 N^{\beta }\big)^{6 b} \beta ^3 \, _2F_1\big(3-9 b,2-\frac{2}{\beta
   };3-\frac{2}{\beta };\frac{G_0 N^{\beta }}{G_1}\big) \big(1-\frac{G_0 N^{\beta
   }}{G_1}\big)^{-9 b}}{G_1^2 (\beta -1)}\\ \nonumber &+\frac{1440 A b G_0^2 N^{2 \beta -2} \big(G_1-G_0
   N^{\beta }\big)^{6 b} \beta ^3 \, _2F_1\big(3-9 b,2-\frac{2}{\beta };3-\frac{2}{\beta };\frac{G_0 N^{\beta
   }}{G_1}\big) \big(1-\frac{G_0 N^{\beta }}{G_1}\big)^{-9 b}}{G_1^2 (\beta -1)}\\ \nonumber &+\frac{26640 A
   b^2 G_0^2 N^{2 \beta -2} \big(G_1-G_0 N^{\beta }\big)^{6 b} \beta ^2 \, _2F_1\big(3-9
   b,2-\frac{2}{\beta };3-\frac{2}{\beta };\frac{G_0 N^{\beta }}{G_1}\big) \big(1-\frac{G_0 N^{\beta
   }}{G_1}\big)^{-9 b}}{G_1^2 (\beta -1)}\\ \nonumber &+\frac{4320 A b G_0^2 N^{2 \beta -2} \big(G_1-G_0
   N^{\beta }\big)^{6 b} \beta ^2 \, _2F_1\big(3-9 b,2-\frac{2}{\beta };3-\frac{2}{\beta };\frac{G_0 N^{\beta
   }}{G_1}\big) \big(1-\frac{G_0 N^{\beta }}{G_1}\big)^{-9 b}}{G_1^2 (\beta -1)}\\ \nonumber &-\frac{5760 A b
   G_0^2 N^{2 \beta -2} \big(G_1-G_0 N^{\beta }\big)^{6 b} \beta  \, _2F_1\big(3-9 b,2-\frac{2}{\beta
   };3-\frac{2}{\beta };\frac{G_0 N^{\beta }}{G_1}\big) \big(1-\frac{G_0 N^{\beta
   }}{G_1}\big)^{-9 b}}{G_1^2 (\beta -1)}\\ \nonumber &+\frac{120960 A b^3 G_0^3 N^{3 \beta -2}
   \big(G_1-G_0 N^{\beta }\big)^{6 b} \beta ^3 \, _2F_1\big(3-9 b,3-\frac{2}{\beta };4-\frac{2}{\beta
   };\frac{G_0 N^{\beta }}{G_1}\big) \big(1-\frac{G_0 N^{\beta }}{G_1}\big)^{-9 b}}{G_1^3
   (3 \beta -2)}\\ \nonumber &-\frac{53280 A b^2 G_0^3 N^{3 \beta -2} \big(G_1-G_0 N^{\beta }\big)^{6 b} \beta ^2 \,
   _2F_1\big(3-9 b,3-\frac{2}{\beta };4-\frac{2}{\beta };\frac{G_0 N^{\beta }}{G_1}\big)
   \big(1-\frac{G_0 N^{\beta }}{G_1}\big)^{-9 b}}{G_1^3 (3 \beta -2)}\\ \nonumber &+\frac{5760 A b G_0^3 N^{3
   \beta -2} \big(G_1-G_0 N^{\beta }\big)^{6 b} \beta  \, _2F_1\big(3-9 b,3-\frac{2}{\beta };4-\frac{2}{\beta
   };\frac{G_0 N^{\beta }}{G_1}\big) \big(1-\frac{G_0 N^{\beta }}{G_1}\big)^{-9 b}}{G_1^3
   (3 \beta -2)}\\ \nonumber &+\frac{2880 A b G_0 N^{\beta -2} \big(G_1-G_0 N^{\beta }\big)^{6 b} \beta ^3 \,
   _2F_1\big(3-9 b,\frac{\beta -2}{\beta };2-\frac{2}{\beta };\frac{G_0 N^{\beta }}{G_1}\big)
   \big(1-\frac{G_0 N^{\beta }}{G_1}\big)^{-9 b}}{G_1 (\beta -2)}\\ \nonumber &-\frac{8640 A b G_0 N^{\beta -2}
   \big(G_1-G_0 N^{\beta }\big)^{6 b} \beta ^2 \, _2F_1\big(3-9 b,\frac{\beta -2}{\beta };2-\frac{2}{\beta
   };\frac{G_0 N^{\beta }}{G_1}\big) \big(1-\frac{G_0 N^{\beta }}{G_1}\big)^{-9 b}}{G_1
   (\beta -2)}\\ \nonumber &+\frac{5760 A b G_0 N^{\beta -2} \big(G_1-G_0 N^{\beta }\big)^{6 b} \beta  \,
   _2F_1\big(3-9 b,\frac{\beta -2}{\beta };2-\frac{2}{\beta };\frac{G_0 N^{\beta }}{G_1}\big)
   \big(1-\frac{G_0 N^{\beta }}{G_1}\big)^{-9 b}}{G_1 (\beta -2)}\\ \nonumber &-\frac{18240 A b^2 G_0^2 N^{2
   \beta -3} \big(G_1-G_0 N^{\beta }\big)^{6 b} \beta ^4 \, _2F_1\big(4-9 b,2-\frac{3}{\beta
   };3-\frac{3}{\beta };\frac{G_0 N^{\beta }}{G_1}\big) \big(1-\frac{G_0 N^{\beta
   }}{G_1}\big)^{-9 b}}{G_1^2 (2 \beta -3)}\\ \nonumber &+\frac{1920 A b G_0^2 N^{2 \beta -3} \big(G_1-G_0
   N^{\beta }\big)^{6 b} \beta ^4 \, _2F_1\big(4-9 b,2-\frac{3}{\beta };3-\frac{3}{\beta };\frac{G_0 N^{\beta
   }}{G_1}\big) \big(1-\frac{G_0 N^{\beta }}{G_1}\big)^{-9 b}}{G_1^2 (2 \beta -3)}\\ \nonumber &+\frac{47520 A
   b^2 G_0^2 N^{2 \beta -3} \big(G_1-G_0 N^{\beta }\big)^{6 b} \beta ^3 \, _2F_1\big(4-9
   b,2-\frac{3}{\beta };3-\frac{3}{\beta };\frac{G_0 N^{\beta }}{G_1}\big) \big(1-\frac{G_0 N^{\beta
   }}{G_1}\big)^{-9 b}}{G_1^2 (2 \beta -3)}\\ \nonumber &-\frac{29280 A b^2 G_0^2 N^{2 \beta -3}
   \big(G_1-G_0 N^{\beta }\big)^{6 b} \beta ^2 \, _2F_1\big(4-9 b,2-\frac{3}{\beta };3-\frac{3}{\beta
   };\frac{G_0 N^{\beta }}{G_1}\big) \big(1-\frac{G_0 N^{\beta }}{G_1}\big)^{-9 b}}{G_1^2
   (2 \beta -3)}\\ \nonumber &-\frac{10560 A b G_0^2 N^{2 \beta -3} \big(G_1-G_0 N^{\beta }\big)^{6 b} \beta ^2 \,
   _2F_1\big(4-9 b,2-\frac{3}{\beta };3-\frac{3}{\beta };\frac{G_0 N^{\beta }}{G_1}\big)
   \big(1-\frac{G_0 N^{\beta }}{G_1}\big)^{-9 b}}{G_1^2 (2 \beta -3)}\\ \nonumber &+\frac{8640 A b G_0^2 N^{2
   \beta -3} \big(G_1-G_0 N^{\beta }\big)^{6 b} \beta  \, _2F_1\big(4-9 b,2-\frac{3}{\beta };3-\frac{3}{\beta
   };\frac{G_0 N^{\beta }}{G_1}\big) \big(1-\frac{G_0 N^{\beta }}{G_1}\big)^{-9 b}}{G_1^2
   (2 \beta -3)}\\ \nonumber &+\frac{32640 A b^3 G_0^3 N^{3 \beta -3} \big(G_1-G_0 N^{\beta }\big)^{6 b} \beta ^4 \,
   _2F_1\big(4-9 b,3-\frac{3}{\beta };4-\frac{3}{\beta };\frac{G_0 N^{\beta }}{G_1}\big)
   \big(1-\frac{G_0 N^{\beta }}{G_1}\big)^{-9 b}}{G_1^3 (\beta -1)}\\ \nonumber &-\frac{3680 A b^2 G_0^3 N^{3
   \beta -3} \big(G_1-G_0 N^{\beta }\big)^{6 b} \beta ^4 \, _2F_1\big(4-9 b,3-\frac{3}{\beta
   };4-\frac{3}{\beta };\frac{G_0 N^{\beta }}{G_1}\big) \big(1-\frac{G_0 N^{\beta
   }}{G_1}\big)^{-9 b}}{G_1^3 (\beta -1)}\\ \nonumber &+\frac{160 A b G_0^3 N^{3 \beta -3} \big(G_1-G_0
   N^{\beta }\big)^{6 b} \beta ^4 \, _2F_1\big(4-9 b,3-\frac{3}{\beta };4-\frac{3}{\beta };\frac{G_0 N^{\beta
   }}{G_1}\big) \big(1-\frac{G_0 N^{\beta }}{G_1}\big)^{-9 b}}{G_1^3 (\beta -1)}\\ \nonumber &-\frac{32640 A
   b^3 G_0^3 N^{3 \beta -3} \big(G_1-G_0 N^{\beta }\big)^{6 b} \beta ^3 \, _2F_1\big(4-9
   b,3-\frac{3}{\beta };4-\frac{3}{\beta };\frac{G_0 N^{\beta }}{G_1}\big) \big(1-\frac{G_0 N^{\beta
   }}{G_1}\big)^{-9 b}}{G_1^3 (\beta -1)}
\end{align}

\begin{align}
    \\ \nonumber &-\frac{15840 A b^2 G_0^3 N^{3 \beta -3}
   \big(G_1-G_0 N^{\beta }\big)^{6 b} \beta ^3 \, _2F_1\big(4-9 b,3-\frac{3}{\beta };4-\frac{3}{\beta
   };\frac{G_0 N^{\beta }}{G_1}\big) \big(1-\frac{G_0 N^{\beta }}{G_1}\big)^{-9 b}}{G_1^3
   (\beta -1)}\\ \nonumber &+\frac{960 A b G_0^3 N^{3 \beta -3} \big(G_1-G_0 N^{\beta }\big)^{6 b} \beta ^3 \,
   _2F_1\big(4-9 b,3-\frac{3}{\beta };4-\frac{3}{\beta };\frac{G_0 N^{\beta }}{G_1}\big)
   \big(1-\frac{G_0 N^{\beta }}{G_1}\big)^{-9 b}}{G_1^3 (\beta -1)}\\ \nonumber &+\frac{19520 A b^2 G_0^3 N^{3
   \beta -3} \big(G_1-G_0 N^{\beta }\big)^{6 b} \beta ^2 \, _2F_1\big(4-9 b,3-\frac{3}{\beta
   };4-\frac{3}{\beta };\frac{G_0 N^{\beta }}{G_1}\big) \big(1-\frac{G_0 N^{\beta
   }}{G_1}\big)^{-9 b}}{G_1^3 (\beta -1)}\\ \nonumber &+\frac{1760 A b G_0^3 N^{3 \beta -3} \big(G_1-G_0
   N^{\beta }\big)^{6 b} \beta ^2 \, _2F_1\big(4-9 b,3-\frac{3}{\beta };4-\frac{3}{\beta };\frac{G_0 N^{\beta
   }}{G_1}\big) \big(1-\frac{G_0 N^{\beta }}{G_1}\big)^{-9 b}}{G_1^3 (\beta -1)}\\ \nonumber &-\frac{2880 A b
   G_0^3 N^{3 \beta -3} \big(G_1-G_0 N^{\beta }\big)^{6 b} \beta  \, _2F_1\big(4-9 b,3-\frac{3}{\beta
   };4-\frac{3}{\beta };\frac{G_0 N^{\beta }}{G_1}\big) \big(1-\frac{G_0 N^{\beta
   }}{G_1}\big)^{-9 b}}{G_1^3 (\beta -1)}\\ \nonumber &-\frac{107520 A b^4 G_0^4 N^{4 \beta -3}
   \big(G_1-G_0 N^{\beta }\big)^{6 b} \beta ^4 \, _2F_1\big(4-9 b,4-\frac{3}{\beta };5-\frac{3}{\beta
   };\frac{G_0 N^{\beta }}{G_1}\big) \big(1-\frac{G_0 N^{\beta }}{G_1}\big)^{-9 b}}{G_1^4
   (4 \beta -3)}\\ \nonumber &+\frac{97920 A b^3 G_0^4 N^{4 \beta -3} \big(G_1-G_0 N^{\beta }\big)^{6 b} \beta ^3 \,
   _2F_1\big(4-9 b,4-\frac{3}{\beta };5-\frac{3}{\beta };\frac{G_0 N^{\beta }}{G_1}\big)
   \big(1-\frac{G_0 N^{\beta }}{G_1}\big)^{-9 b}}{G_1^4 (4 \beta -3)}\\ \nonumber &-\frac{29280 A b^2 G_0^4 N^{4
   \beta -3} \big(G_1-G_0 N^{\beta }\big)^{6 b} \beta ^2 \, _2F_1\big(4-9 b,4-\frac{3}{\beta
   };5-\frac{3}{\beta };\frac{G_0 N^{\beta }}{G_1}\big) \big(1-\frac{G_0 N^{\beta
   }}{G_1}\big)^{-9 b}}{G_1^4 (4 \beta -3)}\\ \nonumber &+\frac{2880 A b G_0^4 N^{4 \beta -3} \big(G_1-G_0
   N^{\beta }\big)^{6 b} \beta  \, _2F_1\big(4-9 b,4-\frac{3}{\beta };5-\frac{3}{\beta };\frac{G_0 N^{\beta
   }}{G_1}\big) \big(1-\frac{G_0 N^{\beta }}{G_1}\big)^{-9 b}}{G_1^4 (4 \beta -3)}\\ \nonumber &-\frac{17280 A
   G_0 N^{\beta +1} \big(G_1-G_0 N^{\beta }\big)^{6 b} \, _2F_1\big(4-9 b,1+\frac{1}{\beta
   };2+\frac{1}{\beta };\frac{G_0 N^{\beta }}{G_1}\big) \big(1-\frac{G_0 N^{\beta
   }}{G_1}\big)^{-9 b}}{G_1 (\beta +1)}\\ \nonumber &+\frac{25920 A G_0^2 N^{2 \beta +1} \big(G_1-G_0
   N^{\beta }\big)^{6 b} \, _2F_1\big(4-9 b,2+\frac{1}{\beta };3+\frac{1}{\beta };\frac{G_0 N^{\beta
   }}{G_1}\big) \big(1-\frac{G_0 N^{\beta }}{G_1}\big)^{-9 b}}{G_1^2 (2 \beta +1)}\\ \nonumber&-\frac{17280 A
   G_0^3 N^{3 \beta +1} \big(G_1-G_0 N^{\beta }\big)^{6 b} \, _2F_1\big(4-9 b,3+\frac{1}{\beta
   };4+\frac{1}{\beta };\frac{G_0 N^{\beta }}{G_1}\big) \big(1-\frac{G_0 N^{\beta
   }}{G_1}\big)^{-9 b}}{G_1^3 (3 \beta +1)}\\ \nonumber &+\frac{4320 A G_0^4 N^{4 \beta +1} \big(G_1-G_0
   N^{\beta }\big)^{6 b} \, _2F_1\big(4-9 b,4+\frac{1}{\beta };5+\frac{1}{\beta };\frac{G_0 N^{\beta
   }}{G_1}\big) \big(1-\frac{G_0 N^{\beta }}{G_1}\big)^{-9 b}}{G_1^4 (4 \beta +1)}\\ \nonumber &+4320 A n
   \big(G_1-G_0 N^{\beta }\big)^{6 b} \, _2F_1\big(4-9 b,\frac{1}{\beta };1+\frac{1}{\beta };\frac{G_0
   N^{\beta }}{G_1}\big) \big(1-\frac{G_0 N^{\beta }}{G_1}\big)^{-9 b}+\frac{5 b G_0 N^{\beta -1} \beta
    \, _2F_1\big(2-3 b,\frac{\beta -1}{\beta };2-\frac{1}{\beta };\frac{G_0 N^{\beta }}{G_1}\big)
   \big(1-\frac{G_0 N^{\beta }}{G_1}\big)^{-3 b}}{G_1 (\beta -1)}\\ \nonumber &+\frac{480 A b G_0
   N^{\beta -3} \big(G_1-G_0 N^{\beta }\big)^{6 b} \beta ^4 \, _2F_1\big(4-9 b,\frac{\beta -3}{\beta
   };2-\frac{3}{\beta };\frac{G_0 N^{\beta }}{G_1}\big) \big(1-\frac{G_0 N^{\beta
   }}{G_1}\big)^{-9 b}}{G_1 (\beta -3)}\\ \nonumber &-\frac{2880 A b G_0 N^{\beta -3} \big(G_1-G_0
   N^{\beta }\big)^{6 b} \beta ^3 \, _2F_1\big(4-9 b,\frac{\beta -3}{\beta };2-\frac{3}{\beta };\frac{G_0 N^{\beta
   }}{G_1}\big) \big(1-\frac{G_0 N^{\beta }}{G_1}\big)^{-9 b}}{G_1 (\beta -3)}\\ \nonumber &+\frac{5280 A b
   G_0 N^{\beta -3} \big(G_1-G_0 N^{\beta }\big)^{6 b} \beta ^2 \, _2F_1\big(4-9 b,\frac{\beta
   -3}{\beta };2-\frac{3}{\beta };\frac{G_0 N^{\beta }}{G_1}\big) \big(1-\frac{G_0 N^{\beta
   }}{G_1}\big)^{-9 b}}{G_1 (\beta -3)}\\ \nonumber &-\frac{2880 A b G_0 N^{\beta -3} \big(G_1-G_0
   N^{\beta }\big)^{6 b} \beta  \, _2F_1\big(4-9 b,\frac{\beta -3}{\beta };2-\frac{3}{\beta };\frac{G_0 N^{\beta
   }}{G_1}\big) \big(1-\frac{G_0 N^{\beta }}{G_1}\big)^{-9 b}}{G_1 (\beta -3)}-84 A G
   \big(G_1-G_0 N^{\beta }\big)^{2 b} \big(1-\frac{G_0 N^{\beta }}{G_1}\big)^{-5 b}-5 \big(1-\frac{G_0 N^{\beta
   }}{G_1}\big)^{-3 b}\\ \nonumber &+\frac{240 A
   b^2 G G_0^2 N^{2 \beta -1} \big(G_1-G_0 N^{\beta }\big)^{2 b} \beta ^2 \, _2F_1\big(2-5
   b,2-\frac{1}{\beta };3-\frac{1}{\beta };\frac{G_0 N^{\beta }}{G_1}\big) \big(1-\frac{G_0 N^{\beta
   }}{G_1}\big)^{-5 b}}{G_1^2 (2 \beta -1)}\\ \nonumber &-\frac{60 A b G G_0^2 N^{2 \beta -1} \big(G_1-G_0
   N^{\beta }\big)^{2 b} \beta  \, _2F_1\big(2-5 b,2-\frac{1}{\beta };3-\frac{1}{\beta };\frac{G_0 N^{\beta
   }}{G_1}\big) \big(1-\frac{G_0 N^{\beta }}{G_1}\big)^{-5 b}}{G_1^2 (2 \beta -1)}\\ \nonumber &-\frac{60 A b
   G G_0 N^{\beta -1} \big(G_1-G_0 N^{\beta }\big)^{2 b} \beta ^2 \, _2F_1\big(2-5 b,\frac{\beta
   -1}{\beta };2-\frac{1}{\beta };\frac{G_0 N^{\beta }}{G_1}\big) \big(1-\frac{G_0 N^{\beta
   }}{G_1}\big)^{-5 b}}{G_1 (\beta -1)}\\ \nonumber &+\frac{60 A b G G_0 N^{\beta -1} \big(G_1-G_0
   N^{\beta }\big)^{2 b} \beta  \, _2F_1\big(2-5 b,\frac{\beta -1}{\beta };2-\frac{1}{\beta };\frac{G_0 N^{\beta
   }}{G_1}\big) \big(1-\frac{G_0 N^{\beta }}{G_1}\big)^{-5 b}}{G_1 (\beta -1)}-\frac{5 b G_0 N^{\beta -1}
   \beta ^2 \, _2F_1\big(2-3 b,\frac{\beta -1}{\beta };2-\frac{1}{\beta };\frac{G_0 N^{\beta }}{G_1}\big)
   \big(1-\frac{G_0 N^{\beta }}{G_1}\big)^{-3 b}}{G_1 (\beta -1)}\\ \nonumber &+180 A G n
   \big(G_1-G_0 N^{\beta }\big)^{2 b} \, _2F_1\big(-5 b,\frac{1}{\beta };1+\frac{1}{\beta };\frac{G_0
   N^{\beta }}{G_1}\big) \big(1-\frac{G_0 N^{\beta }}{G_1}\big)^{-5 b}+\frac{10 b^2 G_0^2 N^{2
   \beta -1} \beta ^2 \, _2F_1\big(2-3 b,2-\frac{1}{\beta };3-\frac{1}{\beta };\frac{G_0 N^{\beta }}{G_1}\big)
   \big(1-\frac{G_0 N^{\beta }}{G_1}\big)^{-3 b}}{G_1^2 (2 \beta -1)}\\ \nonumber &-\frac{5 b G_0^2 N^{2 \beta
   -1} \beta  \, _2F_1\big(2-3 b,2-\frac{1}{\beta };3-\frac{1}{\beta };\frac{G_0 N^{\beta }}{G_1}\big)
   \big(1-\frac{G_0 N^{\beta }}{G_1}\big)^{-3 b}}{G_1^2 (2 \beta -1)}+84 A G \big(G_1-G_0 N^{\beta }\big)^{2 b}+2880 A \big(G_1-G_0
   N^{\beta }\big)^{6 b}\Biggl]
   \label{eqb1}
\end{align}
}%

\end{document}